\documentstyle[10pt,epsf]{article}

\title{Identifying nonlinear wave interactions
	in plasmas \\ using 
	two-point measurements: \\
	A case study of Short Large 
	Amplitude \\ Magnetic Structures (SLAMS)} 

\author{\begin{tabular}{c}  
T. Dudok de Wit$^1$, V. V. Krasnosel'skikh$^2$,
	M. Dunlop$^3$, 
	H. L\"uhr$^4$
\\
\\
{\small $^1$ Centre de Physique Th\'eorique, CNRS and 
	Universit\'e de Provence, Marseille, France}\\
{\small $^2$ Laboratoire de Physique et Chimie
	de l'Environnement, CNRS, Orl\'eans, France}\\
{\small $^3$ Space and Atmospheric Physics, 
	Imperial College, London, United Kingdom}\\
{\small $^4$ GeoForschungsZentrum, 
	Potsdam, Germany}\\
\\
{\small 27 February 1999}
\\
{\small to appear in J. Geoph. Res.}
\end{tabular}}

\date{}

\def\grl{{\it Geophys. Res. Lett.}}
\def\jgr{{\it J. Geoph. Res.}}
\def\om{\omega}

\begin{document}  
\maketitle

\begin{abstract}
Two fundamental quantities for characterizing nonlinear
wave phenomena in plasmas are the spectral energy transfer
associated with the energy redistribution between
Fourier modes, and the linear growth rate.
It is shown how these quantities can be estimated
simultaneously from dual-spacecraft data using Volterra 
series models. We consider magnetic field data gathered upstream 
the Earth's quasiparallel bow shock, in which Short Large 
Amplitude Magnetic Structures (SLAMS) supposedly play a leading 
role. The analysis attests the dynamic evolution of the SLAMS
and reveals an energy cascade toward high-frequency waves. These
results put constraints on possible mechanisms for the shock 
front formation.
\end{abstract}


\section{Introduction}
\label{sec:intro}

In this paper, a Volterra series representation
is used to describe the nonlinear evolution in time 
and in space of a fluctuating wave field. The basis for this
approach is that plasmas 
can often be viewed as a causal nonlinear system
(a ``black box'')
that reacts to a given excitation by giving
a response. By modeling the nonlinear transfer function
associated with this system,
deeper insight can be gained into the underlying
physics. 

The analysis of the nonlinear transfer function is detailed
here for the particular
case where two-point measurements are available.
First, we show how to model the dynamical response. 
Then, the physical 
interpretation of the model coefficients is given.
Particular attention is paid to the linear growth rate, 
which expresses the linear instability of the wave field, 
and to the spectral 
energy transfer, which describes how the instabilities 
saturate through nonlinear wave interactions.

We apply this method to
magnetic field data gathered by the dual AMPTE
satellites near the Earth's quasiparallel bow shock.
This data set corresponds to a regime of quasi-stationary 
turbulence in a collisionless plasma;
it has received much interest in relation with
the existence of Short Large Amplitude Magnetic 
Structures (SLAMS) [{\it Schwartz et al.,} 1992].
It is shown how a transfer function analysis
reveals the role played by these nonlinear structures.

This paper is divided in three parts.
The experimental context is described in 
section~\ref{sec:context}.
Sections~\ref{sec:model} to \ref{sec:choice} 
are devoted to data analysis aspects with a 
description of the model, the choice of its parameters, and its 
validation. Finally, in
sections~\ref{sec:linear} and \ref{sec:nonlinear}, experimental
data are analyzed and interpreted.


\section{The Experimental Context}
\label{sec:context}

\begin{figure}[!htb]
\centerline{\epsfbox{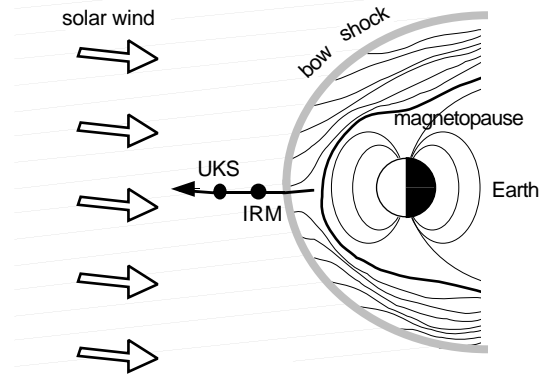}}
\caption{Configuration of the Earth's bow shock showing
the magnetic field lines, the orientation of the solar wind, and
the orbit of the spacecraft for the event considered in 
this paper.}
\label{fig:bowshock}
\end{figure}

The magnetic field data of interest were gathered 
by the dual Active Magnetospheric Particle Tracer
Explorers spacecraft (United Kingdom Satellite 
(AMPTE-UKS) and Ion Release Module (AMPTE-IRM))  on day 
304 of 1984 just 
upstream the Earth's quasi-parallel bow shock. Several 
studies have already been devoted to this particular event 
[{\it Schwartz and Burgess,} 1991, {\it Schwartz et al.,} 1992; 
{\it Mann et al.,}
1994; {\it Dudok de Wit and Krasnosel'skikh,} 1995;
{\it Dudok de Wit et al.,} 1995],
which provides a paradigm for
nonlinear effects in turbulence. The
spacecraft were closely following each other on the
same outbound orbit (with a separation of $\delta x=144$ km),
depicted in Figure\ \ref{fig:bowshock}.

A distinctive feature of the studied region is the 
occurrence of SLAMS, which
supposedly play a leading role in the shock front
formation [{\it Schwartz et al.,} 1992]. 
The shock wave is caused by the sudden
deceleration of the supersonic solar wind at the 
encounter of the Earth's magnetosphere. The SLAMS grow
out of low-frequency waves that propagate away
from the shock front but are convected back toward
the Earth by the solar wind [{\it Thomsen et al.}, 1990]. 
This steepening process is likely to result from an interaction 
with ion beams coming from the shock 
front [{\it Scholer}, 1993].

There are several open questions regarding the
role played by SLAMS. Quasi-parallel shocks are currently
viewed either as an entity [{\it Winske et al.}, 1990]
or as a patchy transition zone made by a merging of SLAMS 
[{\it Schwartz and Burgess,} 1991]. The relationship
between the SLAMS and the whistler wave packets that frequently 
occur at their leading
edge is not well understood either, although there is
numerical [{\it Omidi and Winske,} 1990] and experimental
[{\it Dudok de Wit and Krasnosel'skikh,} 1995] evidence for
a causal link between the two.

\begin{figure}[!htb]
\centerline{\epsfbox{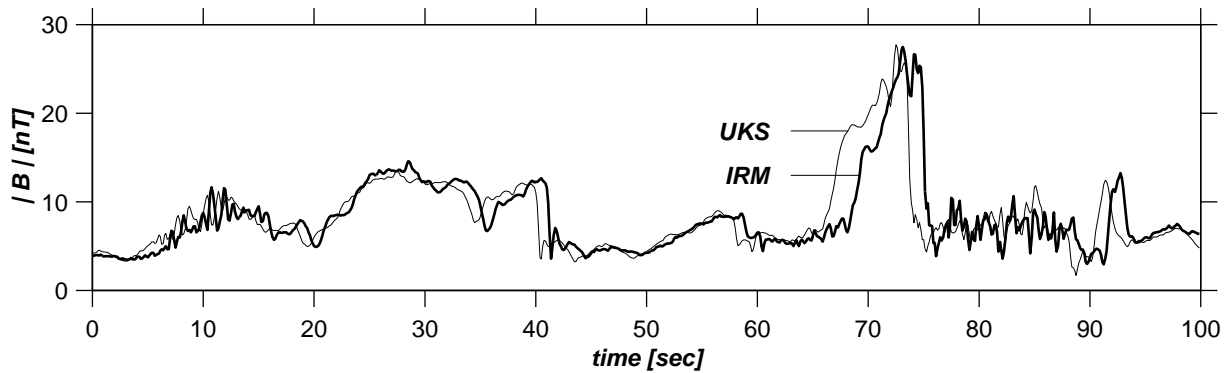}}
\caption{Excerpt of the time evolution of the magnetic field
amplitude, as measured by the two spacecraft. 
A typical SLAMS appears at $t=$71 s. The precursor whistler
wave appears at its trailing edge since the wave field is
convected backward by the strong solar wind.}
\label{fig:raw}
\end{figure}

An excerpt of the magnetometer data is shown in
Figure\ \ref{fig:raw}.
The trajectory of the spacecraft, the prevailing
magnetic field and the average solar wind velocity 
($v_{sw}=$ 370 km/s) are all
parallel within a few degrees. This is an important
point since it means that both spacecraft see the same structures,
separated by a time interval of about 1 s.
A comparative analysis should therefore
reveal how the wave field, and in particular the SLAMS,
evolve as they move from one spacecraft to the other.
We do this by building a Volterra model that tries to predict
the wave field of AMPTE-IRM using the data of 
AMPTE-UKS as input.

Each spacecraft provides a data set which consists of
the three components 
of the magnetic field, measured immediately upstream the 
shock front. For each component the number of samples 
is $4521$; the data were sampled at a 
constant rate of 8 Hz after being low-pass filtered at 4 Hz.
We have chosen to consider the three components as different
ensembles, thereby artificially increasing the sample size
by a factor of 3.
The anisotropy of the wave field a priori does not justify
such an approximation, but  no
significant differences were found between the model coefficients 
as estimated separately from each component.
An obvious future extension would be to have a model that
takes into account the vectorial nature of the wave field.

\begin{figure}[!htb]
\centerline{\epsfbox{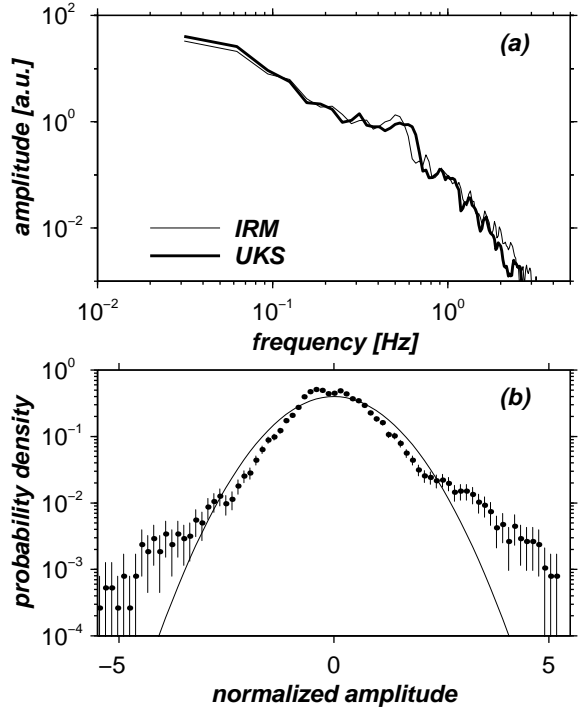}}
\caption{(a) Power spectral density and (b) 
probability distribution of the magnetic 
field of AMPTE-UKS projected along the direction of 
maximum variance. The solid line corresponds to a
Gaussian distribution with the same variance;
all amplitudes are normalized to have unit variance.}
\label{fig:spectra}
\end{figure}

The power spectral density of the wave field is
illustrated in Figure\ \ref{fig:spectra}a and can be
qualified as being continuous and essentially featureless. 
Notice that all frequencies are expressed 
in the spacecraft reference
frame, in which they are Doppler-shifted by the strong
solar wind.
The spectral densities are almost the same
for the two spacecraft. Figure\ \ref{fig:spectra}b shows 
the wave field probability distribution, which has 
non-Gaussian tails. The departure from 
Gaussianity should be underlined, since it is
a necessary condition for having nonlinear wave-wave 
interactions [{\it Kim and Powers,} 1979].

\section{Modeling the Nonlinear Transfer Function}
\label{sec:model}

Much work has been done on the theory of nonlinear transfer
functions in turbulence
[e.g., {\it Monin and Yaglom}, 1975; {\it Krommes}, 1997]
but relatively
little is known about their inference from experimental data,
which can be an unwieldy task. Early results were obtained
in the context of neutral fluid 
turbulence 
[{\it Uberoi,} 1963; {\it Van Atta and Chen,} 
1969; {\it Lii et al.,} 1982; {\it Ritz et al.,} 1988a]
and later in plasmas 
[{\it Ritz and Powers,} 1986; {\it Ritz et al.,}
1988b; {\it Ritz et al.,} 1989; {\it Kim et al.,} 1996]. Powers, 
Ritz and their coworkers contributed to the development of 
a computational framework for two-point measurements 
[{\it Ritz and Powers,} 1986; {\it Ritz et al.,} 1989], 
thereby rendering the 
technique easily accessible to a large class of experiments. 
Their results, however, have remained overlooked, presumably
because of the apparent computational investment and the 
difficulty in validating estimates that are prone to errors. 
In this paper we show how to increase the robustness of the
estimates by using continuous
wavelet transforms instead of the usual Fourier transform.

\subsection{Volterra Series}

Consider a stationary wave field which is measured in time
and in space, and let $y(x,t)$ denote fluctuations around 
a fixed value. We are interested in 
describing the dynamics of this wave field
with the following general model:
\begin{equation}
\frac{\partial{y(x,t)}}{\partial x} = F \big( y(x,t) \big) \ ,
\end{equation}
where $F(y)$ is a continuous, nonlinear and 
time-invariant operator.
{\it Wiener} [1958] showed that for
a large class of causal systems $F(y)$ can be expanded as
a Volterra (or Volterra-Fr\'echet-Wiener) series 
[{\it Schetzen,} 1980], which we write here after taking
the Fourier transform of the time variable
\begin{eqnarray}
\label{eq:volterra2}
\lefteqn{\frac{\partial{y(x,\om)}}{\partial x} = 
		\Gamma(\om) \: y(x,\om)} \\
	&+& \int \Gamma(\om_1,\om_2) \: y(x,\om_1) 
		\: y(x,\om_2)  \nonumber \\
	& & \hspace{7mm} \times \ \delta(\om_1+\om_2-\om) 
		\; d \om_1 d \om_2  \nonumber \\
	&+& \int\!\!\!\int \Gamma(\om_1,\om_2,\om_3) 
		\: y(x,\om_1) \: y(x,\om_2)  \nonumber \\
	& & \hspace{7mm} \times \ y(x,\om_3) 
		\: \delta(\om_1+\om_2+\om_3-\om) 
		\: d \om_1 d \om_2 
			d \om_3 \nonumber \\
	&+& \cdots \nonumber 
\end{eqnarray}
The kernels $\Gamma$ are directly related to the 
higher-order spectra of the process and have a physical meaning. 
$\Gamma(\om)$, $\Gamma(\om_1,\om_2)$, and 
$\Gamma(\om_1,\om_2,\om_3)$ are respectively called the 
linear, quadratic, and cubic interaction terms. 
The generic situation corresponds to
a leading linear term, which
describes the linear dynamics of the system, 
such as the linear growth rate and the dispersion. The 
quadratic term expresses three-wave processes in which 
interactions occur within triads of waves that satisfy the 
resonance condition
\begin{equation}
\label{eq:3omega}
\om = \om_1 + \om_2 \ .
\end{equation}
The cubic term similarly describes four-wave 
processes whose frequencies satisfy the selection rules
\begin{equation}
\label{eq:4omega}
\om = \om_1 + \om_2 + \om_3 
\ \ \mbox{\rm or} \ \
\om + \om_1 = \om_2 + \om_3 \ .
\end{equation}
The main motivation for using a Volterra series expansion stems
from its ability to describe various weakly nonlinear processes in
plasmas [{\it Kadomtsev,} 1982],
ranging from generic drift wave turbulence 
[{\it Balk et al.,} 1990; {\it Horton and Hasegawa,} 1994] 
to Langmuir turbulence as described by the Zakharov equations 
[{\it Musher et al.,} 1995]. Particular attention has been given
to Hamiltonian systems [{\it Zakharov et al.,} 1985] in which
the kernels can be calculated explicitly. The resonant interactions 
defined by (\ref{eq:3omega}) and (\ref{eq:4omega}) are further 
known to be the 
building elements of turbulence as observed in collisionless
plasmas: the decay and modulational 
instabilities, for example, are adequately described in terms
of three-wave and four-wave interactions 
[{\it Krasnosel'skikh and Lefeuvre,} 1993].

Theoretical and experimental 
considerations show that for weak turbulence the
low-order Volterra kernels are the predominant ones. 
Indeed, the characteristic timescale 
associated with the action of a $q$th-order
kernel increases with $q$, making low-order kernels 
much more likely to rule the dynamics
[{\it Zakharov et al.,} 1985]. 
In practice, (\ref{eq:volterra2}) may thus
safely be truncated after the cubic term and quite
often even a quadratically nonlinear model suffices.

\subsection{Strong Versus Weak Turbulence}

The nonlinear model of (\ref{eq:volterra2}) formally 
applies to weak tur\-bu\-lence only, in which the dispersion
and the characteristic growth rates of the Fourier modes
are small. Solar wind turbulence, on the other hand, is
often considered as being of the strong turbulence type. The region
we study is actually a mixture between the two since the
dynamical properties of the wave field are dominated by
a small population of energetic ions interacting with a
plasma of the weak turbulence type.
The weakness of the dispersion 
[{\it Dudok de Wit et al.,}, 1995] and the 
relatively small value of the linear growth
rate (see Section~\ref{sec:linear}) support the validity
of the weak turbulence approximation here.

The extension from weak to
strong turbulence as a first approximation implies
a  loosening of 
the resonance conditions 
(equations (\ref{eq:3omega})-(\ref{eq:3omegak})) 
to account for the finite bandwidth 
of the wave packets [{\it Horton and Hasegawa,} 1994]. 
We shall take this spectral broadening implicitly
into account by
projecting the wave field on wavelets instead of
Fourier modes (see Section~\ref{sec:wavelet}).

\subsection{Spatial Versus Temporal Description}

Equation (\ref{eq:volterra2}) is actually a particular case of a 
class of models 
that describe both the spatial and the temporal structure of 
the wave field. In a more general setting, both 
wavenumbers and  frequencies must satisfy resonance
conditions. Energy and momentum conservation
force three-wave interactions to
occur along the resonant manifold
\begin{equation}
\label{eq:3omegak}
\om({\mathbf k}_1 + {\mathbf k}_2) = \om_1({\mathbf k}_1)
	+ \om_2({\mathbf k}_2) \ ,
\end{equation}
where ${\mathbf k}$ denotes the wave-number vector. 

The wavenumber
dependence of the interaction is often omitted 
by lack of spatially resolved experiments. It 
raises an important point, which is the
separation between spatial and temporal scales, and
the distinction between stationarity and homogeneity.
In our experiment, the wave field is convected
past the satellites by the solar wind and so
the angular frequency $\om_{sat}$ 
we observe in the spacecraft frame is in fact Doppler-shifted,
giving, $\om_{sat} = \om_{pl} + 
		{\mathbf k} \cdot {\mathbf v}_{sw}$.
However, because the wavenumber ${\mathbf k}$ and the
solar wind velocity ${\mathbf v}_{sw}$ are almost
parallel, and because of the fast solar wind, we
may approximate
$\om_{sat} \approx \om_{pl} + 
		k \, v_{sw} \approx k \, v_{sw}$.
This expression
suggests that the spatial structure of the wave field,
projected on the solar wind velocity vector, 
can be probed simply by measuring the time evolution 
and viceversa. This approximation is  
known as the Taylor hypothesis and allows us
to exchange temporal dynamics and spatial
structure
\begin{equation}
\label{eq:Taylor}
	\frac{\partial y}{\partial t} \longleftrightarrow 
		v_{sw} \frac{\partial y}{\partial x} \ .
\end{equation}

\begin{figure}[!htb]
\centerline{\epsfbox{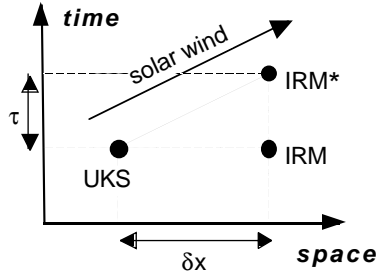}}
\caption{Representation of the spacecraft in space-time,
showing the correspondence between spatial separation 
$\delta x$ and 
time delay $\tau$. The actual position of IRM must be moved
to IRM$^*$ to compensate for the effect of the solar wind.}
\label{fig:spacetime}
\end{figure}

There now remains to convert the Eulerian representation
of the experiment into a Lagrangian one,
in which the magnetic structures are followed from one spacecraft
to the other. A space-time representation of
the spacecraft (see Figure\ \ref{fig:spacetime})
indeed reveals that by taking the difference of the
spacecraft signals, we mix the wave field time derivative 
and the spatial gradient.
A Galilean transformation is needed
$y(x,t) \rightarrow y(x,t'=t-x/v_{sw})$,
which we do by shifting the AMPTE-IRM time series 
by $\tau = -\delta x / v_{sw} = -0.39$ s. 
An additional correction of $-0.28$ s is needed to 
compensate for differences in timing conventions 
[see {\it Schwartz et al.}, 1992].

\subsection{Inferring the Model Coefficients}

Physical insight into our model
can be gained by introducing the real-valued 
density of waves $E(\om,x)$, by analogy to the number of 
quasi particles in condensed matter theory. For a large
ensemble of waves with different frequencies, the
random-phase approximation holds, giving
\begin{equation}
\label{eq:random_phase}
\langle y(\om_1,x) y^*(\om_2,x) \rangle = 
	E(\om_1,x) \delta(\om_1-\om_2) \ .
\end{equation}
Angle brackets denote ensemble-averaging, which is often replaced
by time averaging, assuming ergodicity. 
From equations (\ref{eq:volterra2}) and (\ref{eq:random_phase}) 
we obtain the kinetic equation
\begin{eqnarray}
\label{eq:kinetic}
\lefteqn{\frac{\partial{E(\om,x)}}{\partial t} = 
	2 \overline{\gamma}(\om) E(\om,x)}  \\
	&+& 2 \: \int\!\!\int  T(\om_1,\om_2) 
		\; \delta(\om_1+\om_2-\om)
		\; d \om_1 \: d \om_2
		\nonumber \\
	&+& \cdots \nonumber 
\end{eqnarray}
which models the nonlinear evolution taking place 
between the two spacecraft. Notice that we used the
Taylor hypothesis (equation (\ref{eq:Taylor})) 
to interchange spatial
and temporal derivatives. The quantities of interest are
the average linear growth rate in time
\begin{equation}
	\overline{\gamma}(\om) = v_{sw} \: 
	{\mathrm Re} \left[ \Gamma(\om) \right]
\end{equation}
and the average quadratic energy transfer rate
\begin{eqnarray}
\label{eq:transfer}
\lefteqn{T(\om_1,\om_2) = v_{sw} \: {\mathrm Re} 
	\left[ \: \Gamma(\om_1,\om_2) \right. } \\
	&\times& \left. \langle y(\om_1,x) \: y(\om_2,x) \: 
		y^*(\om_1+\om_2,x) \rangle \: \right] \ . \nonumber
\end{eqnarray}
The latter
attests the existence of non­local interactions which
are a hallmark of nonlinearity. Equation (\ref{eq:kinetic})
shows that the energy flux in Fourier space
$\partial{E} / \partial{t}$ results from a
balance between energy dissipation (or gain) at a
given frequency $\om$ and spectral energy transfers 
between $\om$ and other frequencies.

Nonlinear transfer functions have the advantage
of revealing both the magnitude and the orientation
of spectral energy fluxes: positive values of 
$T(\om_1,\om_2)$ correspond to three-wave interactions 
in which spectral components with angular frequencies $\om_1>0$
and $\om_2>0$ transfer energy to the component
$\om = \om_1 + \om_2$. We shall write this as
$\om_1 + \om_2 \rightarrow \om$. Conversely, negative 
values correspond to decay processes 
$\om \rightarrow \om_1 + \om_2$.

There is a close resemblance between the definition of
the energy transfer function (equation (\ref{eq:transfer}))
and that of the auto bispectrum
[{\it Mendel,} 1991],
\begin{equation}
\label{eq:bispectrum}
B(\om_1,\om_2) = \langle y(\om_1,x) \: y(\om_2,x) 
	\: y^*(\om_1+\om_2,x) \rangle \ ,
\end{equation}
which has been widely used for quantifying
quad\-ratic wave interactions in plasmas 
[{\it Kim and Powers,} 1979;
{\it Lagoutte et al.,} 1989; {\it LaBelle and Lund,} 1992;
{\it P\'ecseli et al.,} 1993;
{\it Dudok de Wit and Krasnosel'skikh,} 1995; 
{\it Bale et al.,} 1996]. 
Transfer functions, however, are more informative
since they detect the presence of
nonlinear interactions
between the observation points irrespective of what happened
farther upstream. Consider for example a 
wave field that underwent nonlinear interactions during its
early history but is now fully static.
This wave field will have a nonzero cross-bispectrum 
even though nonlinear interactions aren't 
actually taking place. 
A nonzero bispectrum thus does not necessarily 
attest the existence of wave-wave interactions
at the observation point. 
Such a caveat was put forward in an analytic example 
by {\it P\'ecseli and Trulsen} [1993]. 
The energy transfer function on the contrary detects
whether energy is being exchanged between spectral modes,
causing the amplitudes and the phases
to vary locally in time and in space. This new information
can be accessed only by comparing the wave field
as it goes from one observation point to the other.
Finally, we note that the existence of energy transfers
presupposes a weak nonstationarity or inhomogeneity of the
wave field.

\subsection{Symmetries}

The real-valued nature of the data and the definition of the
transfer function automatically give rise to a number of
symmetry relations, which
shrink the principal domain in frequency space,  
significantly reducing the number of
model coefficients to be computed 
[see {\it Nam and Powers,} 1994].
The principal domain of the energy transfer
function is even smaller, since 
for $\om_1 + \om_2 = \om$ we have
\begin{equation}
\label{eq:symmetry}
	T(\om_1,\om_2) = T(\om_2,\om_1) =
	-T(\om,-\om_1) = -T(\om,-\om_2) \ .
\end{equation}
%


\section {Estimating the Nonlinear Transfer Function}
\label{sec:estimation}

Our principal problem is the robust estimation of 
Volterra kernels from finite and noise-corrupted data.
This  problem may be alleviated 
by assuming a Gaussian probability 
distribution of the wave field, since the
different regressors can then be identified separately.
This assumption, however, rarely holds in practice. 
Incidentally, it is precisely the nonlinearity that 
causes the distribution to depart from Gaussianity. 
We therefore follow a more general procedure along
the line developed by 
{\it Ritz and Powers} [1986] and 
later improved by {\it Kim and Powers,} [1988]. 

For discrete values of the frequency and with two-point
measurements, (\ref{eq:volterra2}) becomes
%
%
%
{\setlength\arraycolsep{2pt}
\begin{eqnarray}
\label{eq:model1}
	\frac{y_{\om}(x\!+\!\delta x) - y_{\om}(x)}{\delta x} &=& 
	\Gamma_{\om} \: y_{\om}(x) \\
	&+&  \frac{1}{2} \! \! \!
	\sum_{\mathop{_{\om_1, \om_2}}\limits_{\om=\om_1+\om_2}}
	 \! \! \!
	\Lambda_{\om_1, \om_2}^{\om} \: y_{\om_1}(x) 
	\: y_{\om_2}(x) \nonumber \\
	&+& \cdots \nonumber 
\end{eqnarray}}
where $y_{\om}(x)$ is the discrete Fourier transform of $y(x,t)$
and $\{\om\}=\{\om_1,\om_2,\cdots,\om_{N_{\om}}\}$ 
are the regularly spaced frequencies.
Without loss of generality we assume that the ensemble
average vanishes $\langle y(x,t) \rangle = 0$.
It is convenient to express the complex wave field 
$y_{\om}(x)$ as
\begin{equation}
\label{eq:complex}
y_{\om}(x) = |y_{\om}(x)| e^{j \phi_{\om}(x)} \ .
\end{equation}
From (\ref{eq:model1}) and (\ref{eq:complex})
and in the limit where $\delta x \rightarrow 0$,
we obtain a new system [{\it Ritz and Powers,} 1986]
\begin{equation}
\label{eq:model2}
Y_{\om} = L_{\om} U_{\om}
	+ \frac{1}{2} 
	\sum_{\mathop{_{\om_1, \om_2}}\limits_{\om=\om_1+\om_2}}
	Q_{\om_1, \om_2}^{\om} U_{\om_1} U_{\om_2}
	+ \cdots
\end{equation}
with 
\begin{eqnarray}
\label{eq:model3}
Y_{\om} &=& y_{\om}(x + \delta x) 		\\
U_{\om} &=& y_{\om}(x) 				\nonumber \\
L_{\om} &=& (\Gamma_{\om} \delta t + 1 - 
	j \: \delta \phi_{\om}) \;
	e^{j \: \delta \phi_{\om}}  		\nonumber \\
Q_{\om_1,\om_2}^{\om} &=& \Lambda_{\om_1,\om_2}^{\om}
	\: \delta t \: e^{j \: \delta \phi_{\om}} \nonumber \\
\delta t &=& \delta x / v_{sw} 			\nonumber \\
\delta \phi_{\om} &=& \phi_{\om}(x+\delta x) - 
	\phi_{\om}(x)  				\nonumber \ .
\end{eqnarray}
From this system, the physical quantities $\Gamma_{\om}$ and
$\Lambda_{\om_1,\om_2}^{\om}$ can be computed directly,
as shown by {\it Ritz et al.} [1989].

Equation (\ref{eq:model2}) formally represents a nonlinear 
transfer function that links an output $Y_{\om}$ 
(the waveform of AMPTE-IRM) to an input $U_{\om}$ 
(the waveform of AMPTE-UKS). 
The estimation of the linear part of such a transfer 
function is a central problem in system identification, for 
which well-established techniques exist 
[{\it Ljung,} 1987; {\it Priestley,} 1981]. 
Comparatively few experimental efforts, however, have 
been directed toward the robust estimation of quadratic and 
higher-order transfer functions 
[{\it Tick,} 1961; {\it Brillinger,} 1970; {\it Billings,} 1980; 
{\it Bendat,} 1990].
For the sake of simplicity, we shall henceforth restrict 
ourselves to quadratically nonlinear
models. 

The simplest solution consists in selecting the model
whose coefficients minimize the squared residual errors 
$\varepsilon_{\om}$ between the measured wave field 
and the predicted one ${\hat Y}_{\om}$
\begin{equation}
\label{eq:model4}
	\varepsilon_{\om} = | Y_{\om} - {\hat Y}_{\om} |^2 \ .
\end{equation}
The problem then reduces to a
multiple linear regression with a unique solution.
For each angular frequency $\om$ we solve 
for ${\mathsf H}_{\om}$,
\begin{equation}
\label{eq:matrix0}
	{\mathsf U}_{\om}{\mathsf H}_{\om} = {\mathsf Y}_{\om} \ ,
\end{equation}
with
%
\begin{eqnarray*}
\label{eq:matrix1}
{\mathsf U}_{\om} &=& \left[ \begin{array}{cccc}
	U_{\om}(1) 	& U_{\om_1}(1) \; U_{\om-\om_1}(1)
			& U_{\om_2}(1) \; U_{\om-\om_2}(1)
			&  \\
	U_{\om}(2) 	& U_{\om_1}(2) \; U_{\om-\om_1}(2)
			& U_{\om_2}(2) \; U_{\om-\om_2}(2)
			& \cdots \\
	\vdots & \vdots & \vdots & \\
	U_{\om}(N_{ens}) & U_{\om_1}(N_{ens}) \; 
				U_{\om-\om1}(N_{ens})
			& U_{\om_2}(N_{ens}) \; 
				U_{\om-\om_2}(N_{ens})
			& 
	\end{array} \right]  \ \ \ \ \ (19)\\
{\mathsf H}_{\om} &=& \left[L_{\om},
		Q_{\om_1,\om-\om_1}^{\om},
		Q_{\om_2,\om-\om_2}^{\om}, \cdots \right]^T \\
{\mathsf Y}_{\om} &=& \left[Y_{\om}(1),Y_{\om}(2), \cdots,
		Y_{\om}(N_{ens}) \right]^T   \ .
\end{eqnarray*}
Numbers refer to different ensembles collected under
identical conditions, ${\footnotesize T}$ denotes transposition
and the number of unknown coefficients is $N_c$.
The conceptual simplicity of this approach and
its straightforward generalization to cubic and 
higher-order interactions are clear advantages.

The nonlinear transfer function can be obtained by solving 
the overdetermined set of equations (equation (\ref{eq:matrix0})) 
using conventional least squares techniques but
deeper insight can be gained by 
multiplying these equations on the left by ${\mathsf U}_{\om}^*$,
giving
%
\begin{displaymath}
\label{eq:matrix3}
\left[ \begin{tabular}{c|c}
	$\langle |U_{\om}|^2 \rangle$ &
	$\langle U_{\om}^* \: U_{\om'} \: U_{\om-\om'}
		\rangle$ \\
	\\ \hline \\
	$\langle U_{\om} \: U_{\om'}^* \: U_{\om-\om'}^*
		 \rangle$ &
	$\; \; \langle U_{\om'}^* \: U_{\om-\om'}^* \:
		U_{\om''} \: U_{\om-\om''} \rangle$ \\
\end{tabular} \right]
{\mathsf H}_{\om}
=
\left[ \begin{array}{c}
	\langle U_{\om}^* \: Y_{\om} \rangle \\ 
	\langle U_{\om'}^* \: U_{\om-\om'}^* 
		\: Y_{\om} \rangle \\ 
	\langle U_{\om''}^* \: U_{\om-\om''}^* 
		\: Y_{\om} \rangle \\ 
	\vdots
\end{array} \right] \ . \ \ \ \ \ (20)
\end{displaymath}
\setcounter{equation}{20}
The leading matrix (also called higher-order autocovariance matrix)
can be divided into four blocks, one
with a second order moment (the power spectral density), two
with third-order moments (the bispectra) and one with 
fourth-order moments.
The fact that moments of various orders are needed to properly
estimate the linear properties of the wave field
recalls the well-known closure problem which is ubiquitous in the 
spectral modelling of turbulence. Equation (20)
also shows how the non-Gaussian nature of the wave field
enters the results. If the wave field were Gaussian, then
the off-diagonal blocks of the higher-order autocovariance
matrix would vanish and a separate estimation of the different
Volterra kernels would be possible.


\section {Wavelet Versus Fourier Transform}
\label{sec:wavelet}

For the solution of (20) to be numerically 
stable and physically relevant, it is essential to have
$N_{ens} \gg N_c$ 
($N_c$ is the number of unknown coefficients)
and ${\mathsf U}_{\om}$ nonsingular.
A compromise is thus needed between $N_{ens}$
and the number 
of different Fourier modes $N_{\om}$, hereafter 
referred to as the mesh resolution.
Usually, time series are divided into 
(possibly overlapping) sequences, each of
which is Fourier transformed. A better compromise
can be achieved with wavelets, which offer
additional resolution in time at the expense of a 
lower frequency resolution. The continuous wavelet transform of 
$y(t)$ is defined as 
\begin{equation}
y(a,\tau) = \int y(t) \frac{1}{\sqrt{a}} \: h^*
	\left( \frac{t-\tau}{a} \right) 
	\: dt \ ,
\end{equation}
where $h(t)$ is the analyzing wavelet and $a$ its scale. The optimum 
tradeoff between time and frequency resolution 
is achieved with Gaussian or Morlet wavelets
\begin{equation}
h(t) = \frac{1}{\pi^{1/4} \: \sigma^{1/2}} 
	e^{2\pi jt} \; e^{-t^2/2\sigma^2} \ ,
\end{equation}
for which each scale is related to an instantaneous angular 
frequency $\om = 2\pi/a$. The frequency resolution, 
defined in terms of the cutoff frequency at 3 dB is
$\Delta\om/\om = 1/4\sigma$ 
and the usual Fourier transform is recovered for 
$\sigma \rightarrow \infty$.

Compared 
to windowed Fourier transforms, the wavelet transforms 
yield statistically better behaved estimates of the spectral 
properties [{\it van Milligen et al.,} 1995; {\it Dudok de Wit
and Krasnosel'skikh,} 1995]. Our motivation, however, is not just 
computational but also stems from the ability of wavelets to 
resolve transient and soliton-like features
[{\it Farge et al.,} 1996].
Indeed, the strongly turbulent magnetic field 
shows transient structures
that are more akin to wavelets than to coherent waves with
an infinite extension.

The main drawback of this approach is 
its greater computational burden, since the number 
of ensembles $N_{ens}$ now almost equals the number of samples.
Furthermore, we are left with a free parameter,
the wavelet width $\sigma$. 
Since a fixed mesh resolution is wanted,
with no spectral overlap between adjacent components
$Y_{\om_j}$ and $Y_{\om_{j+1}}$,
we adapt the wavelet width to the frequency in 
order to have $\sigma \geq \om/4\delta\om$.
An additional condition $\sigma \geq 1$ is imposed to
prevent the analyzing wavelet from being too much distorted.


\section {Validation Criteria for the Transfer Function Model}
\label{sec:validation}

Validation is a key issue in 
Volterra model identification. There exists no single 
satisfactory criterion for performing such a validation,
but to a large extent we can rely 
on well-established techniques that have been 
developed for linear systems [{\it Ljung,} 1987]. 

Since our problem involves the solution of a linear system
of equations, a good starting point is an inspection
of the degree of independence between the 
columns of the matrix ${\mathsf U}_{\om}$ 
(equation (19))
and the output ${\mathsf Y}_{\om}$. The 
correlation function between ${\mathsf Y}_{\om}$ 
and the first column of
${\mathsf U}_{\om}$ 
\begin{equation}
\label{eq:correl_lin}
\gamma_L^2(\om) = 
	\frac{|\langle Y_{\om} U_{\om}^* \rangle|^2}
		{\langle | Y_{\om} |^2 \rangle
		\langle | U_{\om} |^2 \rangle}
\end{equation}
indicates how well the linear transfer function succeeds in 
predicting the output. This is the coherence function,
which is bounded between 0 and 1. Likewise, the 
correlation function between the output and other columns 
of the matrix gives
\begin{equation}
\label{eq:bicoherence}
\gamma_Q^2(\om_1,\om_2) = 
	\frac{|\langle Y_{\om_1+\om_2} U_{\om_1}^* U_{\om_2}^* \rangle|^2}
	{\langle | Y_{\om_1+\om_2} |^2 \rangle
	\langle | U_{\om_1} U_{\om_2} |^2 \rangle} \ .
\end{equation}
This is the cross-bicoherence,
i.e., the cross-bispectrum normalized 
to the power spectral density. 
Its value is bounded between zero for 
uncorrelated waves and unity for triads of waves whose 
phases are totally correlated. For a cubic model, one similarly 
defines the cross-tricoherence, which quantifies the 
strength of four-wave interactions
\begin{equation}
\label{eq:tricoherence}
\gamma_C^2(\om_1,\om_2,\om_3) = 
	\frac{|\langle Y_{\om_1+\om_2+\om_3} 
		U_{\om_1}^* U_{\om_2}^* U_{\om_3}^* \rangle|^2}
	{\langle | Y_{\om_1+\om_2+\om_3} |^2 \rangle
	\langle | U_{\om_1} U_{\om_2} U_{\om_3} |^2 \rangle} \ .
\end{equation}
Note that there exist variants of these definitions 
[e.g. {\it Kravtchenko-Berejnoi et al.,} 1995].

Higher-order coherence 
functions reveal which spectral 
components are likely to be involved in 
nonlinear interactions. They do not, however, 
tells us whether the model is actually good in predicting
the wave field. 
A high bicoherence, for example, does 
not yet justify the choice of a quadratically nonlinear model.
A more global figure of merit is obtained by
comparing the measured output signal $Y_{\om}$   
to the model prediction $\hat{Y}_{\om}$ 
\begin{equation}
\label{eq:correl_cross}
\gamma_{Y \hat{Y}}^2(\om) = 
	\frac{|\langle Y_{\om} \hat{Y}_{\om}^* \rangle|^2}
	{\langle | Y_{\om} |^2 \rangle
		\langle | \hat{Y}_{\om} |^2 \rangle} \ .
\end{equation}
In practice, one half of the data is used to estimate the transfer 
function while the other half is kept for cross-validation. 
These simple prescription tools can be complemented by tests on
the residuals, etc.


\section{Choosing the Right Model}
\label{sec:choice}

The choice of the model parameters involves
three main issues. More specific aspects 
are deferred to the appendix.

\subsection{Choice of the Model Order}

The basic question of the model order should ideally be answered
by computing Volterra kernels for various orders and 
truncating the series as soon as they become negligible.
The finite size of the data does not allow this, 
so the question should rather be:
How faithfully does a truncated low-order model
reproduce the observed dynamics ?

\begin{figure}[!hb]
\centerline{\epsfbox{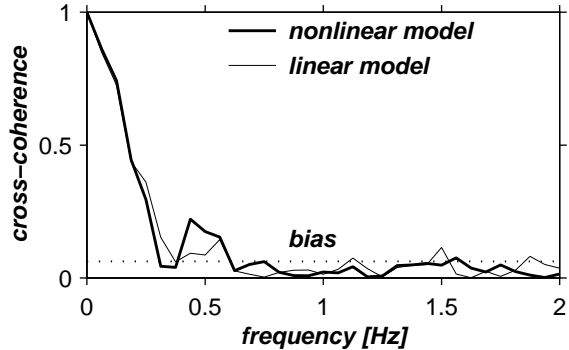}}
\caption{Cross-coherence between the measured and the simulated
output, for a linear and a nonlinear model. Values close to or
below the bias level are not considered to be significant
[{\it Bendat and Piersol,} 1986].}
\label{fig:crosscorr}
\end{figure}

As mentioned before, there are several reasons 
to believe that a low-order model 
should capture most of the dynamics,
especially in weak turbulence. 
This can be verified in different ways. 
{\it Ritz and Powers} [1986] considered nonlinear correlations
between linear and nonlinear terms. We focus instead on
the predictive capacity of the models,
using the cross-validation defined in 
(\ref{eq:correl_cross}). We built first-, second-,
and third-order models, all of which were
tested against the data (the third order 
model could could not have as much frequency resolution
because of its large number of degrees of freedom).

Figure\ \ref{fig:crosscorr}
shows the result of the cross-validation applied
to a linear and to a quadratically nonlinear model. 
Both models succeed relatively well in predicting the low
frequency part of the AMPTE-IRM waveform. The performance
drop with increasing frequency is a well-known effect, 
which cannot be compensated simply by increasing the model 
complexity. Possible causes are
the decreasing signal-to-noise ratio, the finite lifetime
and the dispersion of the wave packets, fluctuations in the
solar wind velocity, and the 1-D approximation of our model.

The central result here is the close performance of the linear
and the quadratic models, which
attests the predominantly linear behavior of the wave field
and a priori supports
the choice of a low-order model. A notable exception occurs 
around $f=\om/2\pi \approx 0.5$ Hz, where a quadratic model
brings some improvement. This shall see later that
nonlinear effects are indeed important in that frequency
band.

\subsection{Choice of the Frequency Range}

There is a strong impetus for reducing as much as possible
the number of degrees of freedom of our model. One way of
doing this is by reducing the frequency range. As shown in 
Figure \ref{fig:crosscorr},
fluctuations with frequencies beyond 0.8 Hz cannot be 
satisfactorily modeled and so one may safely truncate
the frequency range at 1 Hz, above which the power
spectral content becomes negligible anyway. We checked
that higher-order coherence functions vanish as well above 1 Hz.

A further reduction in the number of degrees of freedom can
in principle be achieved by discarding in the linear system 
(equation (\ref{eq:matrix0})) those columns of the matrix which
are not significantly correlated with the output 
${\mathsf Y}_{\om}$. Such a reduction is permitted when
the nonlinear interactions are very localized in
frequency space.

\subsection{Choice of the Mesh Resolution}

The frequency spacing $\delta \om$ (which is 
proportional to $1/N_{\om}$) must be small enough
to distinguish important features such as
spectral lines and yet as large as possible to prevent the model
from being overdetermined.
Since we deal with broadband turbulence, a relatively 
coarse mesh should a priori suffice. Nonlinear
parametric models [{\it Billings,} 1980]
may be needed when closely spaced lines must be resolved.

\begin{figure}[!htb]
\centerline{\epsfbox{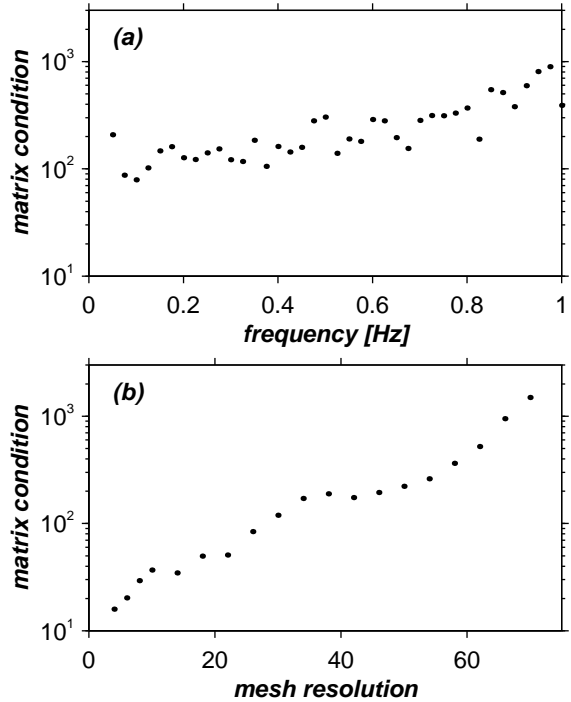}}
\caption{Condition number of the matrix ${\mathsf U}_{\om}$ 
versus (a) frequency with a fixed mesh resolution
$N_{\om}=20$, and (b) versus $N_{\om}$ for a fixed 
frequency $f \approx 0.5$ Hz.}
\label{fig:condition}
\end{figure}

The impact of the mesh resolution is best revealed
by the condition number [{\it Golub and Van Loan,} 1993]
of the matrix ${\mathsf U}_{\om}$, which gives a 
figure of merit for the ill-posedness of (\ref{eq:matrix0}). 
The condition number is at best $1$ and typically should not 
exceed a few hundreds; its value is displayed in 
Figure \ref{fig:condition} for different 
mesh resolutions. The condition 
degrades for increasing $N_{\om}$ because more coefficients
have to be estimated from the same sample; another reason
is the increasing collinearity between 
columns of ${\mathsf U}_{\om}$. 
These problems 
may be partially alleviated by projecting 
${\mathsf U}_{\om}$ on an
orthogonal basis [see {\it Im et al.,} 1996].

From these considerations, we choose a quadratically
nonlinear model with $N_{\om}=40$ and 
a frequency range from 0 to 1 Hz.


\section{Linear Properties of the Wave Field}
\label{sec:linear}

We now focus on the interpretation of the model and
start with the leading term, which is the  linear one.
The linear kernel $\Gamma_{\om}$ can 
conveniently be split into a real and an imaginary part
\begin{equation}
\Gamma(\om) = \overline{\gamma}(\om) + 
	j \overline{\theta}(\om) \ .
\end{equation}
The imaginary part 
\begin{equation}
\label{eq:theta}
	\overline{\theta}(\om) \approx \frac{1}{\delta t} 
	{\mathrm Im} \left[ \log \langle U_{\om}^* \: Y_{\om} 
		\rangle \right] 
\end{equation}
expresses the average 
phase-shift undergone by the wave-packets as they move
from one spacecraft to the other. It is therefore related
to the wavenumber  $\overline{k}$, averaged over the
power spectral density, by 
$\overline{k} = \overline{\theta} / v_{sw}$. 

\begin{figure}[!htb]
\centerline{\epsfbox{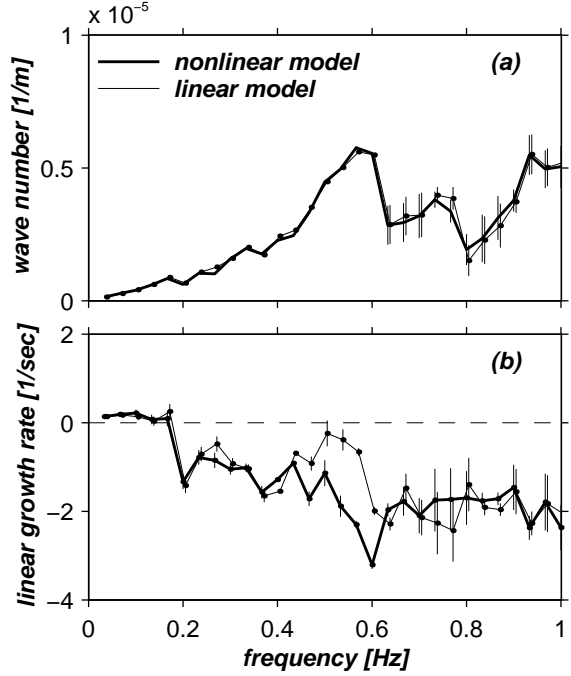}}
\caption{(a) Imaginary and (b) real parts of the
linear transfer function, as calculated using a linear and
a nonlinear model. Error bars correspond to $\pm$ one 
standard deviation. The results are inaccurate above 0.6 Hz
because the model fails to reproduce small-amplitude 
fluctuations correctly.}
\label{fig:linear}
\end{figure}

The average dispersion relation $\overline{k}(\om)$ is shown in 
Figure \ref{fig:linear}.
Given our timing convention,
we are in the plasma rest frame and
positive wavenumbers correspond to a sunward motion.
Error bars correspond to one
standard deviation as calculated from the least-squares
fit of the model. 

Figure\ \ref{fig:linear} shows that the wave field is 
essentially dispersionless up to 
about 0.5 Hz. Above this frequency, the dispersion becomes 
positive, and high-frequency waves move ahead 
of low-frequency ones. We refer to previous work 
[{\it Dudok de Wit et al.,} 1995] for a discussion on this,
but just note that the modeling fails above 0.6 Hz, 
presumably because of the low power spectral density.

The real part of the Volterra kernel 
\begin{equation}
\label{eq:gamma}
	\overline{\gamma}(\om) \approx \frac{1}{\delta t}
	\left(\frac{\left| \langle U_{\om}^* \: Y_{\om} \rangle \right|}
	{\langle U_{\om}^* \: U_{\om} \rangle} - 1 \right)  \ ,
\end{equation}
gives the linear growth averaged as before over $k$. The
results are shown in Figure \ref{fig:linear}. 
The negative value of $\overline{\gamma}(\om)$ attests a
damping of the waves, so we conclude that the wave field
is on average linearly stable. An exception 
occurs below 0.2 Hz, where the wave field 
grows as it goes from one spacecraft to the other. 
Although this growth rate is subject 
to a rather large uncertainty, its positive sign is
statistically significant. Interestingly, this unstable
frequency band coincides with that of the SLAMS and therefore
lends strong support to the instability of these structures.  
The unstable nature of the SLAMS has already been conjectured 
[{\it Schwartz and Burgess,} 1991], but we now have the first 
direct evidence for a dynamic evolution.

From the linear growth rate, one can estimate the 
characteristic time needed for the SLAMS to grow, 
assuming that there is no nonlinear mechanism to
saturate such a growth. We find 
$\tau=1/\overline{\gamma}\approx$ 10 s, 
a value that should be compared to the 
characteristic lifetime of these structures
\begin{equation}
\frac{\tau}{T} = \frac{\overline{\theta}}{\overline{\gamma}} 
	= \frac{{\mathrm Im} (\Gamma)}
        {{\mathrm Re} (\Gamma)} \approx 10\% \ .
\end{equation}
This ratio is sufficiently small to justify a linearization of the 
growth process (and the weak turbulence
approximation) and yet large enough to make the 
instability of the SLAMS easily detectable. 

It is instructive to check here the assumption of nonlinearity
by comparing these results to what we would obtain
by fitting a linear model. The two growth 
rates are compared in
Figure \ref{fig:linear} and a discrepancy appears.
We attribute this to the energy 
redistribution process between Fourier modes, which
is neglected in the linear model and correctly taken into 
account in the nonlinear one. As we shall see later, 
the linear model tries to compensate an energy flux around
0.5 Hz by artificially lowering the damping rate. The use
of a nonlinear model for assessing linear
properties should therefore not be underestimated.


\section{Nonlinear Properties of the Wave Field}
\label{sec:nonlinear}

We now consider the properties of the second-order Volterra
kernel. As mentioned before, this is the only nonlinear
term we can reliably estimate given the amount of data. 

\subsection{Phase Couplings}

The second-order Volterra kernels 
$\Gamma(\om_1,\om_2)$ as such are not very
informative, even though they contain all
the relevant information on the quadratic
couplings. Better insight can be gained by
looking at the the cross-bicoherence
(equation (\ref{eq:bicoherence})), which is indicative of
the strength of the quadratic interactions. In the same way, we
compute the cross-tricoherence (equation (\ref{eq:tricoherence}))
to study cubic interactions, even though
the third-order kernel itself cannot be reliably estimated.

\begin{figure}[!htb]
\centerline{\epsfbox{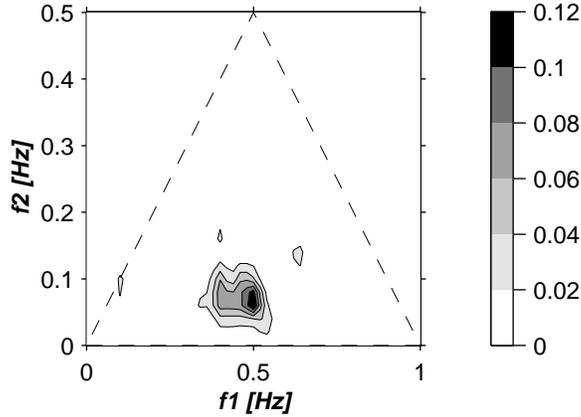}}
\caption{The cross-bicoherence, displayed in the 
principal domain, for frequency-adding 
interactions only ($f_1,f_2 \geq 0$). Its value is 
bounded between 0 and 1.}
\label{fig:bicoh}
\end{figure}

\begin{figure}[!htb]
\centerline{\epsfbox{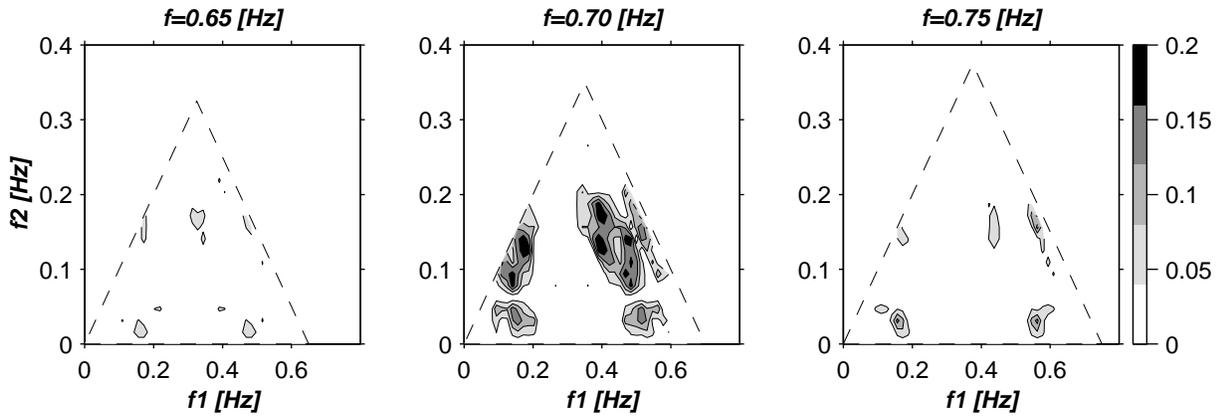}}
\caption{Cross-tricoherence, shown here for interactions 
of the type $f_1+f_2+f_3=f$ Hz only, with
$f_1,f_2,f_3 \geq 0$. Values are bounded between 0 and 1,
and the dashed line refers to the principal domain.}
\label{fig:tricoh}
\end{figure}

The cross-bicoherence and cross-tricoherence 
are displayed in
Figures \ref{fig:bicoh} and 
\ref{fig:tricoh} respectively. 
In both Figures \ref{fig:bicoh} and \ref{fig:tricoh}
the support is restricted to the 
nonredundant and positive frequency domain. 
The most conspicuous result is the presence of 
local maxima that attest the existence of phase couplings 
between specific spectral modes. 
We conclude from the cross-bicoherence that 
a significant phase coupling occurs between wave packets whose
frequencies satisfy the summation rule $0.1 + 0.45 = 0.55$ Hz.
The cross-tricoherence reveals a significant coupling 
for  $0.1 + f_l  + f_m = 0.55$ Hz, with  
$0.1 \leq f_l \leq f_m \leq 0.55$ Hz. Both couplings 
relate wave packets whose frequencies are about 0.1 Hz 
and 0.55 Hz, with possibly some intermediate frequencies 
to enable the phase coupling. As shown in previous work
[{\it Dudok de Wit and Krasnosel'skikh,} 1995],
these characteristic frequencies respectively
correspond to the SLAMS and to the discrete whistler 
wave packets that frequently occur at the leading edge of SLAMS. 
The nonzero cross-bicoherence and cross-tricoherence thus
reveal the existence of a causal relationship between
the SLAMS and the whistlers.

\subsection{Energy Transfers}

\begin{figure}[!htb]
\centerline{\epsfbox{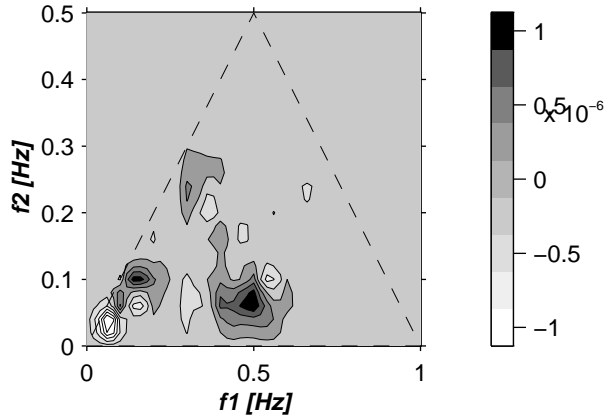}}
\caption{Spectral energy transfer rate (in arbitrary units)
using the same representation as for the cross-bicoherence. The 
confidence interval is about equal to the spacing between
two contour levels.}
\label{fig:transfer}
\end{figure}

The last step now consists in determining whether the SLAMS
and the whistlers are exchanging energy or if they are 
just remnants of a process that took place farther upstream. 
To do so, we compute the
quadratic energy transfer function, shown in
Figure \ref{fig:transfer}.
A significant energy flux appears at
$0.1+0.45 \rightarrow 0.55$ Hz,
which corresponds to an energy transfer going from the SLAMS
to the whistlers. This is the central result of our
paper, from which we conclude that
the whistlers are much more likely to be
a decay product of the SLAMS than some instability
triggered by them. Such a conclusion was partly 
anticipated, but only the energy transfer function
can give unambiguous evidence for it.

The other patterns in Figure \ref{fig:transfer} also
have an interpretation. The $0.1+0.1 \rightarrow 0.2$ Hz 
transfer corresponds a first harmonic generation 
due the nonlinear steepening of the SLAMS.

\subsection{Power Balance}

\begin{figure}[!htb]
\centerline{\epsfbox{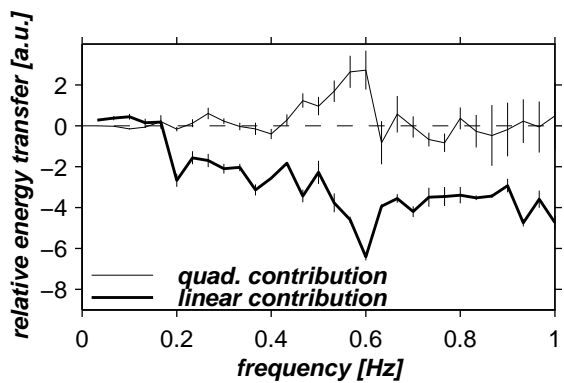}}
\caption{Relative change of spectral power 
showing the contributions
of the linear growth term ($2\overline{\gamma}_{\om}$), 
and of the three-wave interactions
($2\sum T_{\om_1,\om_2} / E_{\om}$). 
The results are unreliable above 0.6 Hz.}
\label{fig:balance}
\end{figure}

Further insight into the dynamics of the wave field
can be gained by studying the power balance. Consider
the truncated second-order 
wave kinetic equation (equation (\ref{eq:kinetic})) 
\begin{equation}
\label{eq:kinetic2}
	\frac{\partial{E_{\om}}}{\partial t} 
	\; = \; 2 \overline{\gamma}_{\om}\,E_{\om} + 2
	\sum_{\stackrel{\om_1,\om_2}{\om=\om_1+\om_2}} 
	T_{\om_1,\om_2}  \ .
\end{equation}
Stationarity ($\partial{E_{\om}}/\partial{t}=0$) 
is approximately reached when the two terms
on the right-hand side cancel; these two terms are
plotted in Figure \ref{fig:balance}.

At low frequencies ($f < $ 0.2 Hz) the growth rate
and the energy transfer indeed approximately cancel
each other and so the wave field
amplitude should not vary much in time. 
We conclude that the decay of the SLAMS is 
approximately compensated by their linear instability.
At higher frequencies the power balance becomes 
increasingly negative, suggesting that the waves are
on average damped. A plausible damping mechanism would be
resonant particle damping, but the various approximations
made in our model may actually cause the damping
to be overestimated at high frequencies (see the appendix).

\subsection{Interpretation}

\begin{figure}[!htb]
\centerline{\epsfbox{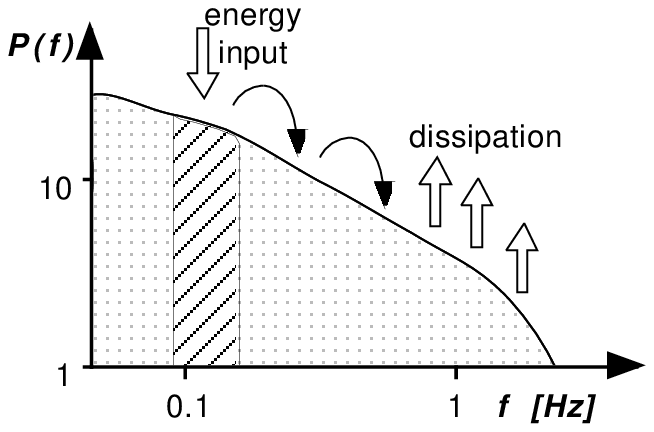}}
\caption{Schematic representation of the
power spectral density of the magnetic field, showing
where the energy enters the wave field and where it is
transferred before being dissipated. The SLAMS are located
in the hatched zone.}
\label{fig:model}
\end{figure}

A coherent scenario now emerges, which
is schematized in Figure \ref{fig:model}. 
The SLAMS appear
as dynamically evolving structures that progressively
grow out of the wave field by drawing energy from
energetic ions. As they grow, nonlinear
effects enter into play. The transfer function analysis
shows that the dominant process is a nonlinear wave
interaction that compensates the growth by an
energy transfer toward high-frequency whistler waves
(some energy may also go into low-frequency
waves). The whistler waves
in turn move ahead of the SLAMS because of the
positive dispersion and are eventually
damped by dissipation. 

The emergence of such right-handed circularly polarized 
waves out of the left-handed
linearly polarized SLAMS shows some striking similarities 
with the expected behavior of solitary waves 
[{\it Hada et al.,} 1989] and also recalls the 
behavior of shock fronts
in weakly dispersive media [{\it Karpman,} 1975].
All these phenomena have in common a competition
between dispersion and nonlinearity, whose distinctive 
manifestation is the resilience of the shape of
the SLAMS.

\begin{figure}[!htb]
\centerline{\epsfbox{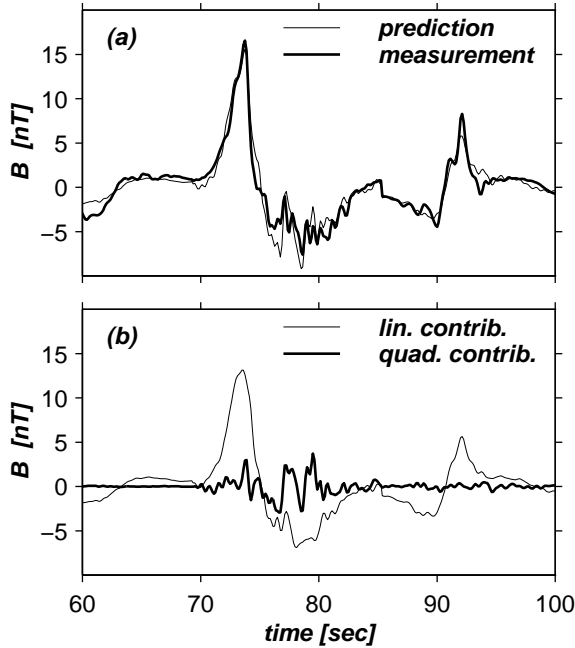}}
\caption{Excerpt of Figure 2, showing
one component of the magnetic field with a typical
SLAMS at $t=$70-76 s and its whistler precursor
at $t=$76-82 s. (a) The wave field
measured by AMPTE-IRM is compared with the model prediction. 
(b) The model prediction is split into its 
linear and quadratic constituents.}
\label{fig:predict}
\end{figure}

To finish, let us 
visualize how the nonlinear interactions show up
in the time domain.
Figure\ \ref{fig:predict} represents a particular
SLAMS with the measured and the predicted wave field. 
We decomposed the latter into its linear 
and quadratic contributions. 
As expected, the linear contribution
captures most of the dynamics but does not correctly reproduce 
the fast oscillations at the leading edge of the SLAMS
and which correspond to a distorted whistler
wave packet. A quadratic contribution is definitely 
needed here to fit the observations.


\section{Conclusions}
\label{sec:conclusions}

This study reveals how Volterra models can be
used to infer nonlinear properties
from a turbulent wave field using two-point measurements.
Provided the model is carefully validated, it
can give direct access to the wave field growth rate and 
to the energy transfer function. 

We used the Volterra approach to analyze plasma turbulence
as observed just upstream the Earth's
quasi-parallel bow shock by the AMPTE spacecraft. An important
feature of the wave field is the occurrence of nonlinear magnetic
structures termed SLAMS. Our analysis attests the
coexistence of two competing mechanisms: the
SLAMS progressively grow by drawing energy from hot ions but
before overturning they saturate and release the excess of energy into high
frequency whistler waves that move ahead of them due to dispersion.
The dynamical evolution of the SLAMS and 
their differential velocity [{\it Schwartz et al.,} 1992]
support the conjecture in which they supposedly 
merge into an extended
front that constitutes the bow shock.

The method we advocate here is  applicable to other types
of events provided they are recorded by multiple-spacecraft
missions whose configuration satisfies some constraints
(see the appendix). A number of improvements can be made, such
as the generalization to vector fields, which would allow 
anisotropy effects to be included. In some cases the addition
of a source term that enforces the stationarity of the
wave field may be desirable.


\subsection*{\bf Acknowledgments}

\bigskip

{\small We would like to thank Steven Schwartz
for making valuable comments on the manuscript. 
Discussions with Arnaud Masson 
on the tricoherence and with Mikhail Balikhin
are also acknowledged.

Michel Blanc thanks Hans P\'ecseli and Vincenzo Carbone
for their assistance in evaluating this paper.}


\appendix
\section{Appendix: Limitations of the Technique}
\label{sec:appendix}

The nonlinear transfer function
cannot be meaningfully assessed without keeping in mind
several limitations and potential pitfalls. The most
important ones are listed here.

\subsection{One-Dimensional Approach} 

Restricting the analysis to two probes means that we can
only study structures propagating parallel to the
spacecraft separation vector. Structures propagating
obliquely to it will cause an overestimation of the
damping rate and an underestimation of the energy
transfer.

\subsection{Anisotropy} 

Our scalar model can in
principle be generalized to vectorial quantities
with the interesting
perspective of addressing anisotropy effects. The price to pay
for this is a larger number of degrees
of freedom. Neglecting the vectorial nature of the
variables may actually alter the power
balance because of the omission of the wave field rotation.
Although the rotation we measure is quite small,
we believe this effect to be sufficient to accentuate the globally
negative trend observed in Figure \ref{fig:balance}.

\subsection{Taylor Hypothesis}

Another problem comes from the connection between 
frequency and wavenumber space, for which the Taylor hypothesis
was invoked at the beginning. 
The dispersionless approximation remains valid up to about
0.5 Hz. Above this limit, the dispersion becomes positive and hence
the resonance conditions 
(equations (\ref{eq:3omega})-(\ref{eq:3omegak})) 
are altered. The finite frequency 
resolution of the wavelets can 
easily accommodate such small detunings in the 
resonance conditions.

\subsection{Validity of the Model}

The weakest point of our approach is the difficulty
in justifying the validity of a low-order model
in a definite way. A quadratically 
nonlinear model suffices for describing
the main features of the wave field
(section \ref{sec:choice}), but Figure\ \ref{fig:tricoh}
reminds us that cubic
interactions cannot be neglected. Although we are confident in
the conclusions drawn from our second-order model,
one should keep in mind that the results
remain approximate as long as cubic and possibly even 
higher-order effects are not taken into account.

\subsection{Calibration of the Probes}

It is essential that the two probes (the 
magnetometers here) be properly calibrated and have the same
instrumental transfer function for our analysis to be
meaningful. Although this is not a problem for the frequency range
we are considering, it may 
exclude diagnostics that have either an
unreliable calibration or a nonlinear response.


\newpage

\section*{References}

\begin{itemize}

\item
Bale, S. D., D.~Burgess, P.~J.~Kellogg, K.~Goetz, R.~L. Howard,
	and S.~J.~Monson,
Evidence of three-wave interactions in the upstream solar wind,
\grl, {\it 23,} 109--112, 1996.

\item
Balk, A. M., V.~E.~Zakharov, and S.~V.~Nazarenko, 
Nonlocal turbulence of drift waves,
{\it Sov. Phys. JETP, 71,} 249--260, 1990.

\item
Bendat, J. S.,
{\it Nonlinear System Analysis and Identification from Random Data,}
John Wiley, New York, 1990.

\item
Bendat, J. S. and A. G. Piersol,
{\it Random Data: Analysis and Measurement Procedures,}
2nd ed.,
John Wiley, New York, 1986.

\item
Billings, S. A.,
Identification of Nonlinear Systems-A Survey,
{\it IEE Proc., Part D, 127,} 272--285, 1980.

\item
Brillinger, D. R.,
The identification of polynomial systems by means of higher
	order spectra,
{\it J. Sound Vibr., 12,} 301--313, 1970.

\item
Dudok de Wit, T., and V.~V.~Krasnosel'skikh, 
Wavelet bicoherence analysis of strong plasma turbulence
	at the Earth's quasiparallel bow shock,
{\it Phys. Plasmas, 2,} 4307--4311, 1995.

\item
Dudok de Wit, T., V.~V.~Krasnosel'skikh, S.~D.~Bale, 
M.~Dunlop, H.~L\"uhr, S.~J.~Schwarz, and L.~J.~C.~Woolliscroft,
Determination of dispersion relations in quasi-stationary
	plasma turbulence using dual satellite data,
\grl, {\it 22,} 2653--2656, 1995.

\item
Farge, M., N.~Kevlahan, V.~Perrier, and E.~Goirand,
Wavelets and turbulence,
{\it Proc. IEEE, 84,} 639--669, 1996.

\item
Golub, G. H., and C.~F.~Van Loan, 
{\it Matrix Computations,} 2nd ed.,
Johns Hopkins Press, Baltimore, Md., 1989.

\item
Hada, T., C.~F.~Kennel, and B.~Buti,
Stationary nonlinear Alfv\'en waves and solitons,
\jgr, {\it 94,} 65--77, 1989.

\item
Horton, W. and A.~Hasegawa, 
Quasi-two-dimensional dynamics of plasmas and fluids
{\it Chaos, 4,} 227--251, 1994.

\item
Im, S., E.~J.~Powers, and I.~S.~Park,
Applications of higher-order statistical signal processing
	to nonlinear phenomena,
in {\it Transport, Chaos and Plasma Physics II,}
edited by S. Benkadda, F. Doveil, and Y. Elskens,
pp. 74--88, World Sci., Singapore, 1996.

\item
Kadomtsev, B.~B.,
{\it Plasma Turbulence,}
Pergamon, New York, 1982.

\item
Karpman,~V.~I.,
{\it Non-linear Waves in Dispersive Media,}
Pergamon, New York, 1975.

\item
Kim, J. S., R. D. Durst, R.~J.~Fonck, E.~Fernandez, 
	A.~Ware, and P.~W.~Terry,
Technique for the experimental estimation of nonlinear energy
	transfer in fully developed turbulence,
{\it Phys. Plasmas, 3,} 3998--4009, 1996.

\item
Kim, K. I., and E.~J.~Powers,
A digital method of modeling quadratically nonlinear systems
	with a general random input,
{\it IEEE Trans. Acoust. Speech Signal Process., ASSP-36,} 
1758--1769, 1988.

\item
Kim, Y. C., and E.~J.~Powers, 
Digital bispectral analysis and its applications to nonlinear
	wave analysis,
{\it IEEE Trans. Plasma Sci., PS-7,} 120--131, 1979.

\item
Krasnosel'skikh, V. V., and F.~Lefeuvre, 
Strong Langmuir turbulence in space plasmas, the problem
	of recognition,
in {\it START: Spatio-temporal analysis for resolving plasma 
turbulence,} Doc. WPP-047, pp. 237--242,
Eur. Space Agency, Paris, 1993.

\item
Kravtchenko-Berejnoi, V., F.~Lefeuvre, V.~V.~Krasno\-sel'skikh, 
	and D.~Lagoutte, 
On the use of tricoherent analysis to detect non-linear
	wave-wave interactions,
{\it Signal Proc., 42,} 291--309, 1995.

\item
Krommes, J. A.,
Systematic statistical theories of plasma turbulence and
intermittency: Current status and future prospects,
{\it Phys. Rep., 283,} 5--48, 1997.

\item
LaBelle, J., and E.~J.~Lund,
Bispectral analysis of equatorial spread $F$ density irregularities,
\jgr, {\it 97,} 8643--8651, 1992.

\item
Lagoutte, D., F.~Lefeuvre, and J.~Hanasz,
Application of bicoherence analysis in study of wave interactions
	in space plasma,
\jgr, {\it 94,} 435--442, 1989.

\item
Lii, K. S., K.~N.~Helland, and M.~Rosenblatt, 
Estimating three-dimensional energy transfer in isotropic
	turbulence,
{\it J. Time Ser. Anal., 3,} 1--28, 1982.

\item
Ljung, L., 
{\it System Identification,}
Prentice-Hall, Englewood Cliffs, N. J., 1987.

\item
Mann, G., H. L\"uhr, and W.~Baumjohann,
Statistical analysis of short large-amplitude magnetic field
	structures in the vicinity of the quasi-parallel bow shock,
\jgr, {\it 99,} 13315--13323, 1994.

\item
Mendel, J. M.,
Tutorial on higher-order statistics (spectra) in signal processing
and system theory: Theoretical results and some applications,
{\it Proc. IEEE, 79,} 278--305, 1991.

\item
Monin, A. S., and A.~M.~Yaglom,
{\it Statistical Fluid Mechanics,} vol. 2, 
MIT Press, Cambridge, Mass., 1975.

\item
Musher, S. L., A. M. Rubenchik, and V.~E.~Zakharov,
Weak Langmuir turbulence,
{\it Phys. Rep., 252,} 177--274, 1995.

\item
Nam, S. W., and E.~J.~Powers,
Application of higher order spectral analysis to cubically
	nonlinear system identification,
{\it IEEE Trans. Acoust. Speech Signal Process., ASSP-42,} 
1746--1765, 1994.

\item
Omidi, N., and D.~Winske, 
Steepening of kinetic magnetosonic waves into shocklets:
simulations and consequences for planetary shocks and comets,
\jgr, {\it 95,} 2281--2300, 1990.

\item
P\'ecseli, H. L., and J.~Trulsen, 
On the interpretation of experimental methods for
investigating nonlinear wave phenomena,
{\it Plasma Phys. Controlled Fusion, 25,} 1701--1715, 1993.

\item
P\'ecseli, H. L., J.~Trulsen, A.~Bahnsen, and F.~Primdahl,
Propagation and nonlinear interaction of low-frequency
	electrostatic waves in the polar cap $E$ region,
\jgr, {\it 98,} 1603--1612, 1993.

\item
Priestley, M. B.,
{\it Spectral Analysis and Time Series,}
Academic Press, San Diego, Calif., 1981.

\item
Ritz, C. P., and E.~J.~Powers, 
Estimation of nonlinear transfer functions for fully
developed turbulence,
{\it Phys. D, 20,} 320--334, 1986.

\item
Ritz, C. P., E.~J.~Powers, R.~W.~Miksad, and R.~S.~Solis, 
Nonlinear spectral dynamics of a transitioning flow,
{\it Phys. Fluids, 31,} 3577--3588, 1988a.

\item
Ritz, C. P., et al.,
Advanced plasma fluctuation analysis techniques and their
	impact on fusion research,
{\it Rev. Sci. Instrum., 59,} 1739--1744, 1988b.

\item
Ritz, Ch. P., E.~J.~Powers, and R.~D.~Bengtson,
Experimental measurement of three-wave coupling and energy cascading,
{\it Phys. Fluids B, 1,} 153--163, 1989.

\item
Schetzen, M.,
{\it The Volterra and Wiener Theories of Nonlinear Systems,}
John Wiley, New York, 1980.

\item
Scholer, M.,
Upstream waves, shocklets, short large-\-amplitude magnetic 
structures and the cyclic behavior of oblique quasi-parallel
collisionless shocks,
\jgr, {\it 98,} 45--57, 1993.

\item
Schwartz, S. J., and D.~Burgess,
Quasi-parallel shocks: A patchwork of three-dimensional structures, 
\grl, {\it 18,} 373--376, 1991.

\item
Schwartz, S. J., D.~Burgess, W.~P.~Wilkinson, R.~L.~Kessel,
	M.~Dunlop, and H.~L\"uhr,
Observations of short large-amplitude magnetic structures
	at a quasi-parallel shock, 
\jgr, {\it 97,} 4209--4227, 1992.

\item
Thomsen, M. F., D.~T.~Gosling, S.~J.~Bame, and C.~T.~Russell,
Magnetic pulsations at a quasi-parallel shock,
\jgr, {\it 95,} 957--966, 1990.

\item
Tick, L. J.,
The estimation of ``transfer functions'' of quadratic systems,
{\it Technometrics, 3,} 563--567, 1961.

\item
Uberoi, M. S.,
Energy transfer in isotropic turbulence,
{\it Phys. Fluids, 6,} 1048--1056, 1963.

\item
van Milligen, B. P., E. S\'anchez, T.~Estrada, C.~Hidalgo,
	B.~Bra\~nas, B.~Carreras, and L.~Garc\'{\i}a,
Wavelet bicoherence: a new turbulence analysis tool,
{\it Phys. Plasmas, 2,} 3017--3032, 1995.

\item
Van Atta, C. W., and W. Y. Chen,
Measurements of spectral energy transfer in grid turbulence,
{\it J. Fluid Mech., 38,} 743--763, 1969.

\item
Wiener, N.,
{\it Nonlinear Problems in Random Theory,}
MIT Press, Cambridge, Mass., 1958.

\item
Winske, D., N.~Omidi, K.~B.~Quest, and V.~Thomas,
Reforming supercritical quasi-parallel shocks, 2,
mechanism for wave generation and front reformation,
\jgr, {\it 95,} 18821--18827, 1990.

\item
Zakharov, V.~E., S.~L.~Musher, and A.~M.~Rubenchik,
Hamiltonian approach to the description of non-linear plasma
	phenomena,
{\it Phys. Rep., 129,} 285--366, 1985.

\end{itemize}

\end{document}